\documentclass[11pt,a4paper]{article}
\pdfoutput=1
\usepackage{jinstpub}

\usepackage{graphicx}
\usepackage{epstopdf}
\usepackage{dcolumn}
\usepackage{bm}
\usepackage{hyperref}

\newcommand{\isot}[2]{$^{\textrm{#2}}$#1 }
\newcommand{\krm}{\isot{Kr}{83m}}

\begin{document}
\title{\boldmath Extraction efficiency of drifting electrons in a two-phase xenon time projection chamber}

\author[,a]{B.~N.~V.~Edwards,$^1$}\footnote{Now at: IBM Research, STFC Daresbury Laboratory, Warrington, WA4 4AD, UK}
\author[a,b,c]{E.~Bernard,}%
\author[a,b,c]{E.~M.~Boulton,}%
\author[,d]{N.~Destefano,$^2$}\footnote{Now at: The MITRE Corporation, 202 Burlington Rd, Bedford, MA 01730}
\author[e]{M.~Gai,}
\author[,a,b]{M.~Horn,$^3$}\footnote{Now at: Sanford Underground Research Facility, 630 E. Summit Street, Lead, SD 57754}
\author[,a]{N.~Larsen,$^4$}\footnote{Now at: University of Chicago, Kavli Institute for Cosmological Physics, 5640 Ellis Ave, Chicago, IL 60637}
\author[,a]{B.~Tennyson,$^5$}\footnote{Now at: H3D, Inc., 3250 Plymouth Rd Suite 203, Ann Arbor, MI 48105}
\author[,a,b,c]{L.~Tvrznikova,$^6$}\footnote{Corresponding author}
\author[,a]{C.~Wahl,$^7$}\footnote{Now at: H3D, Inc., 3250 Plymouth Rd Suite 203, Ann Arbor, MI 48105}%
\author[a,b,c]{D.~N.~McKinsey}%
\affiliation[a]{Yale University, Department of Physics, 217 Prospect St., New Haven, CT 06511, USA}
\affiliation[b]{University of California Berkeley, Department of Physics, Berkeley, CA 94720, USA}
\affiliation[c]{Lawrence Berkeley National Laboratory, Berkeley, CA 94720, USA}
\affiliation[d]{University of Connecticut, Department of Physics, 2152 Hillside Road, Storrs, CT 06269, USA}
\affiliation[e]{University of Connecticut, LNS at Avery Point, 1084 Shennecossett Road, Groton, CT 06340, USA}

\emailAdd{lucie.tvrznikova@yale.edu} 

\abstract{
We present a measurement of the extraction efficiency of quasi-free electrons from the liquid into the gas phase in a two-phase xenon time-projection chamber.  The measurements span a range of electric fields from 2.4 to 7.1~kV/cm in the liquid xenon, corresponding to 4.5 to 13.1~kV/cm in the gaseous xenon.  Extraction efficiency continues to increase at the highest extraction fields, implying that additional charge signal may be attained in two-phase xenon detectors through careful high-voltage engineering of the gate-anode region.
}

\arxivnumber{1710.11032}

\maketitle
\flushbottom

\section{\label{sec:introduction}Introduction}
Two-phase xenon detectors have become a leading detector technology in searches for WIMP (weakly interacting massive particle) dark matter \cite{Akerib:2016,PandaX,XENON100}, searches for axions and axion-like particles \cite{Aprile:2014,Akerib:2017}, and detection of coherent elastic neutrino-nucleus scattering from nuclear reactor, spallation source, and supernova neutrinos \cite{Santos:2011,Akimov:2012, Horowitz:2003, Chakraborty:2014}. They have other potential applications in radiation/particle detection such as searches for neutrinoless double beta decay~\cite{DARWIN,LZTDR} and Compton imaging of gamma-rays \cite{Wahl:2012}.  The understanding of charge and light production in liquid xenon has developed substantially over the past few decades, and the mechanisms which generate detectable signals from energy deposition in the liquid xenon are reviewed in~\cite{henriqueReview, aprileReview}.

The scintillation signal produced by initial atomic excitation and recombining ionization is known as S1, with the light generated from the remaining ionization known as S2.  The generation of S2 signal requires that electrons are drifted upward from the interaction site in the liquid to the liquid surface, where they are extracted into the gas phase. No new light is produced while the electrons drift through the liquid. After extraction into the gas phase, a stronger electric field accelerates the electrons, causing them to produce a proportional electroluminescence signal. 

The size of the S2 signal (in detected photoelectrons) is shown in equation~\ref{eq:S2size}, where $n_i$ is the initial number of ionizations created when the energy is deposited in the interaction, $(1-r)$ is the fraction of the charges not recombining, $\eta$ is the extraction efficiency at the liquid-gas phase boundary, $\xi$ is the electroluminescence yield (photons/electron), $\nu$ is the S2 geometrical light collection (fraction of photons produced that hit a photocathode), $Q$ is the quantum efficiency (fraction of photons hitting the photocathode that generate a photoelectron), and DPE is the double-photoelectron fraction for the electroluminescence \cite{Faham:2015}:

\begin{equation} \label{eq:S2size}
  \textrm{S2 (phe)} = (1-r) \, n_i \, \eta \, \xi \, \, \nu \, Q  (1+DPE).
\end{equation}

In equation~\ref{eq:S2size}, only $\eta$ and $\xi$ are dependent on the electric fields applied in the extraction/electroluminescence region.  Therefore, the extraction efficiency can be determined by varying the extraction field and measuring both the S2 electroluminescent gain for single electrons and the absolute S2 signal size for events of constant energy and drift field. Understanding of the extraction efficiency and how it varies with extraction field is important for reaching optimal sensitivity in experiments using two-phase xenon detector technology.

The potential energy of a free electron in the liquid xenon is lower than in the gaseous phase by 0.67~eV \cite{Tauchert:1975}.  This creates a potential barrier that electrons must cross if they are to be extracted from the liquid into the gas. There are two processes by which this can potentially happen:  (A) thermal emission,  where the tail of the velocity distribution of electrons in thermal equilibrium with the liquid is above the potential barrier, allowing some fraction to be extracted,   and  (B) emission of ``hot'' electrons,  where electrons accelerated by an electric field are imparted with enough energy to overcome the barrier and be extracted. The potential barrier in liquid xenon is much higher than in other condensed noble gases (i.e. argon), and as a result, process (A) provides little contribution to the production of S2 signals from interactions in the liquid xenon, though it is hypothesized to cause delayed single-electron emission on a tens of millisecond time scale \cite{Sorensen:2017}. 

As they drift through the liquid, the electrons are imparted with a non-Maxwellian energy distribution.  The form of this distribution and the amount of energy is dependent on the electric field strength in the liquid \cite{Gushchin:1982,Doke:1981}.  The onset of extraction will not occur until the electrons at the high-energy end of the distribution have enough energy to cross the barrier.  As a result, there is a threshold effect, where an electric field of at least $\sim$~1.5~kV/cm is required to have a non-zero extraction efficiency.  According to the model of \cite{Bolozdynya:1999}, the potential barrier is also reduced as the field is increased, resulting in a lower electron energy required to cross the barrier. The combination of these effects results in a strong electric field dependence of the electron extraction efficiency. 

The extraction efficiency has been measured previously~\cite{Gushchin:1979, Aprile:2014b}. In practice, a given detector is limited in the range of electric fields it can apply to the extraction region.   Here we report a study of relative extraction efficiency over a wide range of fields from 2.4 to 7.1~kV/cm (in the liquid xenon). It is anticipated that these results will be useful in the design optimization of future two-phase xenon experiments. 

\section{\label{sec:pixeyDetector}PIXeY detector}
\begin{figure}[h!]
\begin{center}
\includegraphics[width=0.6\textwidth,clip]{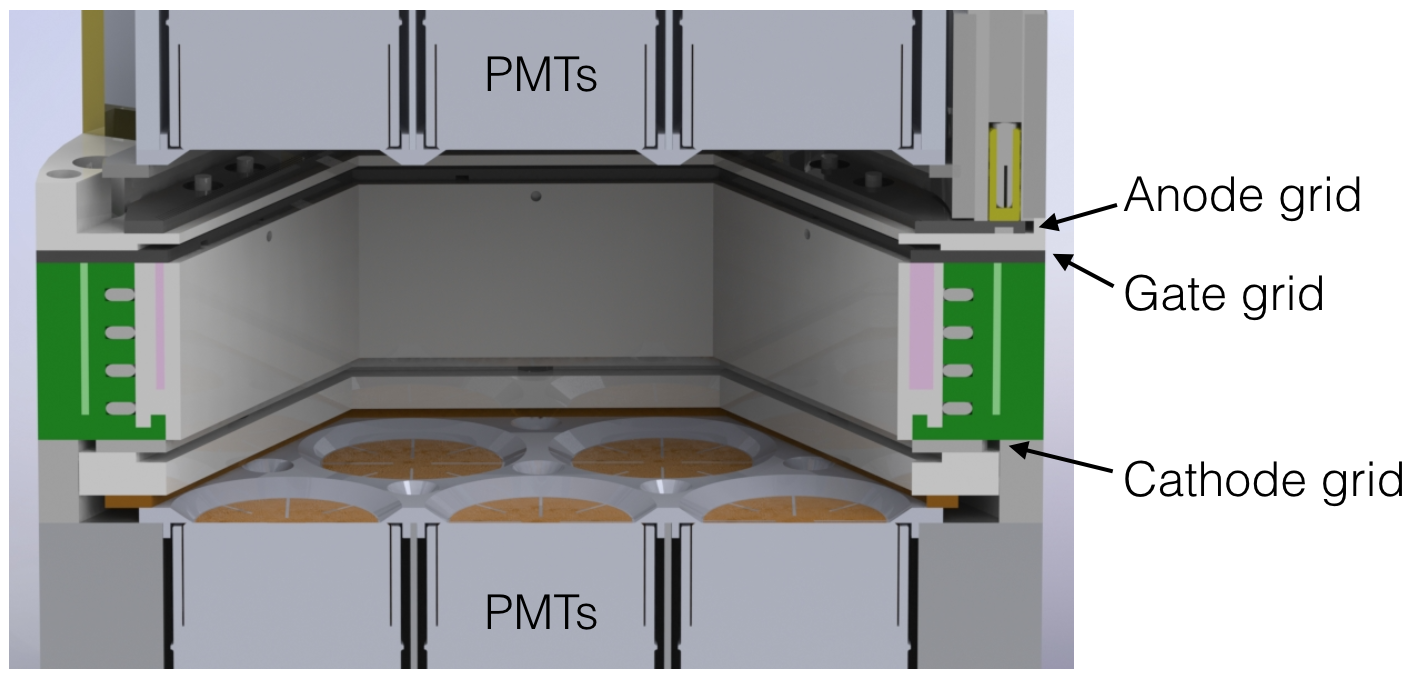}
\caption[]{\label{fig:schematic} Schematic of the PIXeY detector.}
\end{center}
\end{figure}

PIXeY (Particle Identification in Xenon at Yale) is a two-phase xenon detector holding approximately 12~kg of liquid xenon.  The hexagonal time projection chamber (TPC) has an active xenon mass of about 3~kg and is 18.4\,cm in width at its widest point. Figure~\ref{fig:schematic} shows the layout of the PIXeY detector and TPC. The active target region is 5.1~cm deep, defined by the cathode and gate grids.  The anode and gate grids are separated by about 8~mm.

The 175~nm light from S1 and S2 is detected by two arrays of seven Hamamatsu R8778 photomultiplier tubes (PMTs) each, one array above and one below the liquid xenon volume.  The response of the PMTs was normalized using single photoelectron gain measurements.  The PMT response was then confirmed using the data acquired for the study, with an average PMT response of 85~mV-ns per photoelectron, easily resolvable above baseline.  For all data presented in this study, a radial cut of $\sim$5.0~cm is applied using the S2 signal strengths in the top PMT array. 

\begin{figure}[h!]
\begin{center}
\includegraphics[width=0.6\textwidth,clip]{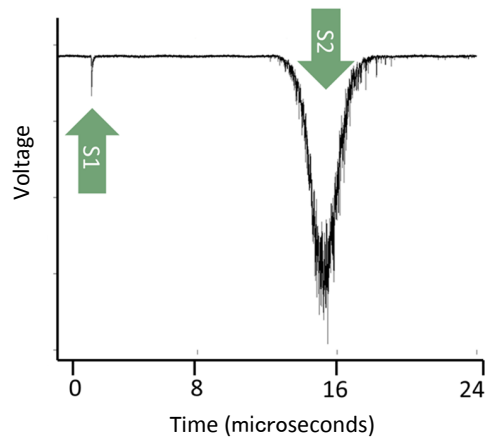}
\caption[]{\label{fig:event_waveform} Example event waveform from the PIXeY detector.}
\end{center}
\end{figure}

The signals from the 14 PMTs underwent $\times8$~amplification before being digitized with a 12-bit ADC (CAEN V1720) waveform digitizer sampled at 250~MHz.  The data acquisition was triggered by the sum signal of the top array PMTs passed through a filter optimized to pass signals with an S2-like timescale. An example event waveform from PIXeY, summed across all 14 PMTs, is shown in figure~\ref{fig:event_waveform}.

\begin{figure}[h!]
\centering
\includegraphics[width=0.6\textwidth]{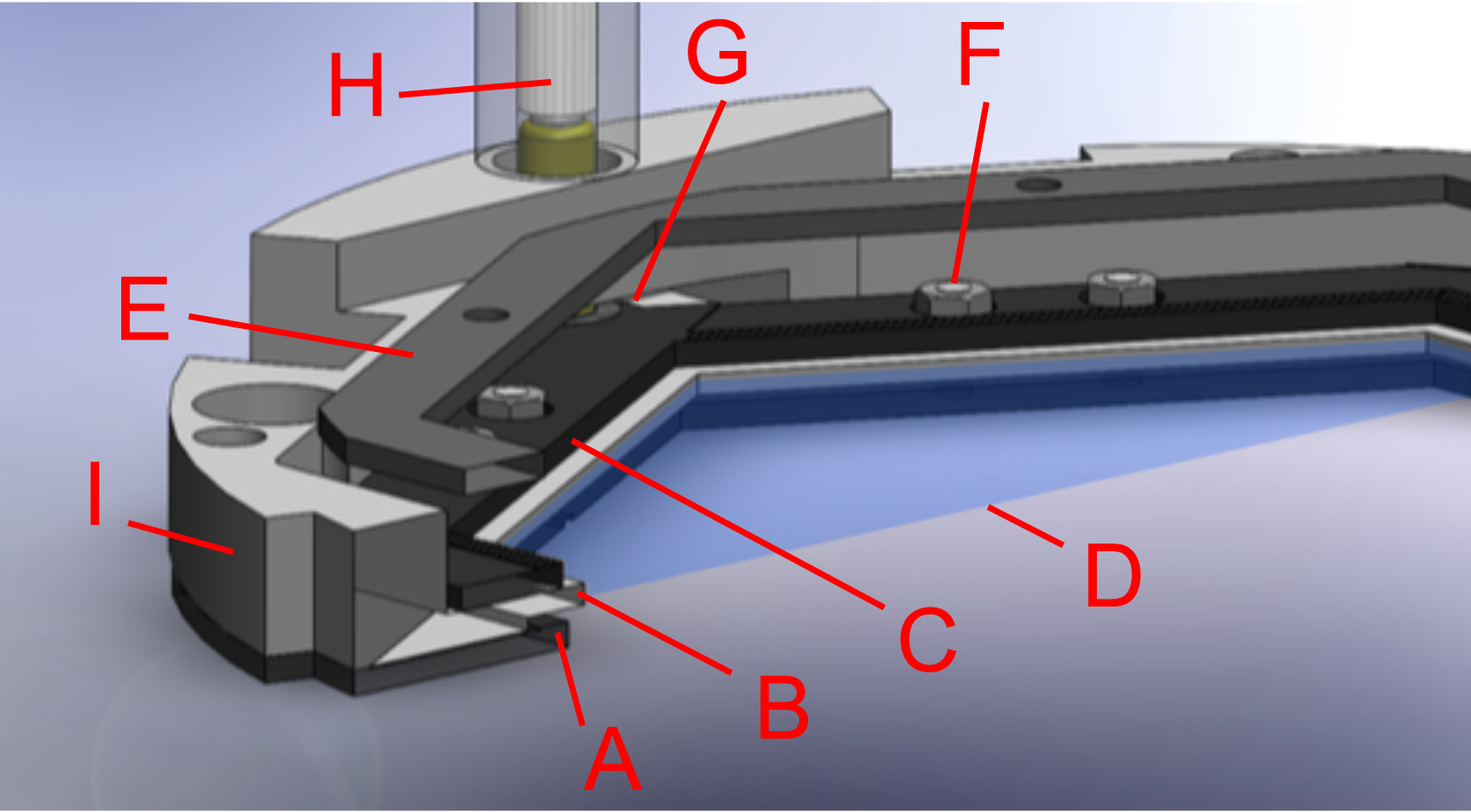}
\caption{Cutaway rendering of the anode region of the TPC.
A - Gate grid frame.
B - Bathtub lip.
C - Anode grid frame.
D - Liquid xenon surface.
E - Top PMT shield grid frame.
F - Anode grid frame retaining posts.
G - Blind pocket with anode HV cable connection.
H - Anode HV cable.
I - Monolithic Teflon bathtub.
Note that features B, F and I are cut from a single contiguous block of Teflon.}
\label{fig:AnodeCut}
\end{figure}

The upper portion of the TPC (shown in figure~\ref{fig:AnodeCut}) was specifically designed to hold high anode voltages, allowing the production of high extraction fields. The anode and gate wire grids are each composed of parallel wires of 80~$\mu$m diameter.  Both grids are soldered at a 1~mm pitch.  The anode grid frame (C) sits within a recessed well cut into a single block of Teflon called the bathtub (I).  The bathtub sits on the gate grid frame (A) and contains a lip (B) that protrudes inward and blocks all direct paths between the anode and gate grid frames.  A weir (not shown) integrated into the bathtub maintains the liquid xenon surface at a constant height (D) coincident with the bathtub lip.  The PMT shield grid frame (E) is suspended above the anode frame by the upper PMT block (not shown).  In this arrangement, the bathtub surfaces exposed to xenon gas contain only long and indirect paths between the anode grid frame and other conductors.  The anode grid frame is secured to the bathtub by Teflon nuts threaded onto retaining posts (F).   One corner of the hexagonal anode grid frame extends into a blind pocket (G) of the bathtub frame.  The pocket contains an electrical receptacle that is accessed by the anode high voltage (HV) cable (H) from above.  The anode HV cable is insulated by polyethylene to 5.8~mm diameter and supplied as model 2149 by Dielectric Sciences.  The upper end of the HV cable terminates at a custom made socket that is sealed by epoxy to a $2\frac{3}{4}$~inch conflat flange.  This allows connection to an external power supply.  The anode grid wires discharged directly to the gate grid wires when this design was tested in argon gas at room temperature.

\section{Results}

\subsection{Electric fields}
The electric fields in PIXeY are defined by the application of high voltage to a series of horizontal grids with parallel wires in the TPC (shown in figure~\ref{fig:schematic}). The ionization yield of drifted electrons (before extraction) is influenced by the electric field within the main drift region, which is set by the voltage applied to the cathode, kept at a constant -1~kV. This produces S2 signals with the same average number of primary electrons drifted from the interaction site, for a given monoenergetic source. The gate grid is always fixed at ground.

The size of the S2 signal seen by the PMTs is determined by a combination of the extraction efficiency (dependent on the field at the liquid surface) and the S2 electroluminescence process (dependent on the field in the gas gap) as defined in equation~\ref{eq:S2size}. Both these fields are set by the voltage applied to the anode grid.

\begin{figure}[h!]
\centering
\includegraphics[width=0.6\textwidth]{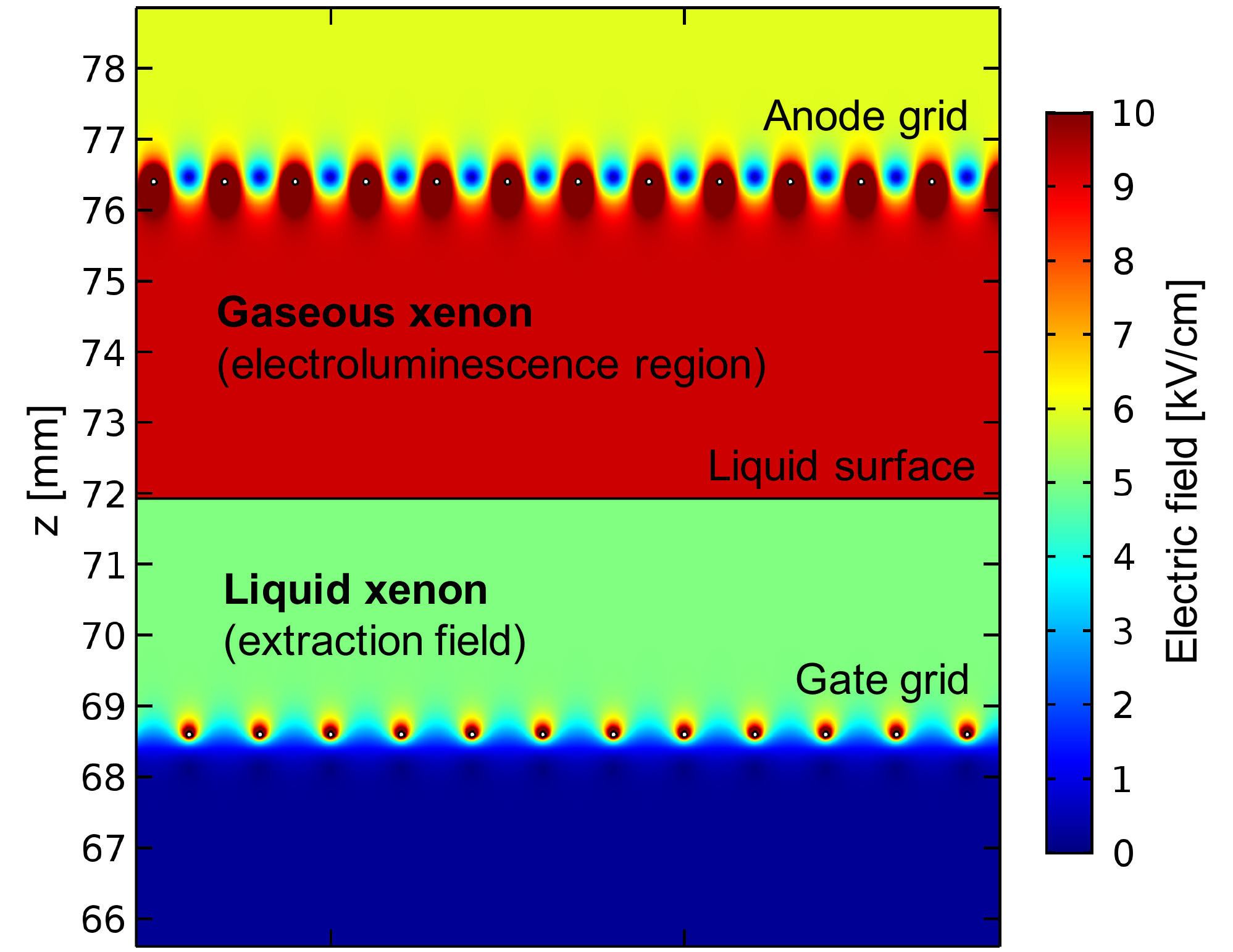}
\caption{Result of a COMSOL electric field simulation of the PIXeY detector.  The region around the liquid-gas phase boundary is shown.  This configuration has a 4.5~mm gas gap (liquid surface to anode) and 6.24~kV applied to the anode grid.}
\label{fig:COMSOLsims}
\end{figure}

A two-dimensional electric field model was developed in COMSOL Multiphysics v5.0\textsuperscript{\textregistered}~\cite{comsol}, a commercially available finite element simulation software, to calculate the electric fields in the different detector regions for each field configuration used. The DC dielectric constant of liquid xenon is assumed to be 1.85, which is typical of measured values in the scientific literature \cite{Amey:1964,Marcoux:1970,Schmidt:2001,Sawada:2003}. Table~\ref{tab:fields} summarizes the modeled voltage configurations, electric fields and systematic uncertainties. Figure~\ref{fig:COMSOLsims} shows the output of one simulation.  The results of these simulations yield extraction fields roughly 12\% lower than would be calculated from a simple parallel-plate model.

The systematic uncertainty in the extraction field comes primarily from the uncertainty in the $z$ separation of the field grids and the location of the gas gap between anode and gate.  The uncertainty in the high voltage applied to the grids is $\sim$ 100~V.  The separation of gate and anode was measured to be $7.8 \pm 0.3$~mm, and the distance from the weir (which sets the liquid height) to the anode was measured as $5.5 \pm 0.3$~mm.  There is some additional uncertainty in the height of the liquid level due to hydrodynamic effects, raising the actual liquid level above the rim of the weir.  This effect has been assessed using the models described in~\cite{Pfister:2013} and estimated to raise the liquid level $1.03 \pm 0.3$~mm above the weir.  Combining these uncertainties, the liquid level is found to be $4.47 \pm 0.52$~mm below the anode grid.  The resulting systematic uncertainties in the electric field are shown in table~\ref{tab:fields}. The extraction field varies negligibly in the $xy$ plane,  within the applied 5~cm radial cut. The change in drift field due to varying extraction field is determined through COMSOL simulations to be less than 18 V/cm for all extraction fields, and this is predicted by the Noble Element Simulation Technique (NEST) \cite{NEST} to produce a sub-1\% change in the number of primary electrons.

\begin{table*}[t]
\caption{Extraction region field configurations studied and their measured relative electron extraction efficiencies, showing the anode voltage applied and the electric fields calculated in liquid and gas phases from a 2D simulation in COMSOL~\cite{comsol}.  The errors shown on the fields are systematic uncertainties dominated by the uncertainty in the $z$ separation of the field grids and $z$ location of the liquid surface.}  \label{tab:fields}
\setlength{\extrarowheight}{2pt}
\begin{center}
\begin{tabular}{| c | c | c | c |}
\hline
{}  & \multicolumn{2}{c|}{Electric fields} & {}\\
Anode & Liquid extraction & Gas electroluminescence & Electron extraction \\
voltage [kV] & region [kV/cm] & region [kV/cm] & efficiency \\
\hline
 3.01 & $ {2.41}\pm{0.12} $ & $ {4.45}\pm{0.22} $ & $ {0.207}\pm{0.006} $\\
 3.23 & $ {2.58}\pm{0.12} $ & $ {4.78}\pm{0.23} $ & $ {0.257}\pm{0.008} $\\
 3.66 & $ {2.93}\pm{0.13} $ & $ {5.42}\pm{0.25} $ & $ {0.361}\pm{0.011} $\\
 4.09 & $ {3.28}\pm{0.14} $ & $ {6.06}\pm{0.27} $ & $ {0.493}\pm{0.015} $\\
 4.30  & $ {3.44}\pm{0.15} $ & $ {6.37}\pm{0.27} $ & $ {0.538}\pm{0.016} $\\
 4.52  & $ {3.62}\pm{0.15} $ & $ {6.70}\pm{0.38} $ & $ {0.568}\pm{0.017} $\\
 4.95  & $ {3.97}\pm{0.16} $ & $ {7.34}\pm{0.30} $ & $ {0.663}\pm{0.020} $\\
 5.38  & $ {4.31}\pm{0.18} $ & $ {7.98}\pm{0.33} $ & $ {0.721}\pm{0.022} $\\
 5.81  & $ {4.66}\pm{0.19} $ & $ {8.62}\pm{0.35} $ & $ {0.790}\pm{0.024} $\\
 6.24  & $ {5.00}\pm{0.20} $ & $ {9.25}\pm{0.37} $ & $ {0.848}\pm{0.025} $\\
 6.45  & $ {5.17}\pm{0.20} $ & $ {9.57}\pm{0.38} $ & $ {0.880}\pm{0.026} $\\
 6.67  & $ {5.35}\pm{0.21} $ & $ {9.89}\pm{0.39} $ & $ {0.883}\pm{0.027} $\\
 6.88  & $ {5.52}\pm{0.22} $ & $ {10.21}\pm{0.40} $ & $ {0.908}\pm{0.027} $\\
 7.10  & $ {5.69}\pm{0.22} $ & $ {10.53}\pm{0.41} $ & $ {0.925}\pm{0.028} $\\
 7.53  & $ {6.04}\pm{0.23} $ & $ {11.17}\pm{0.43} $ & $ {0.954}\pm{0.029} $\\
 8.17  & $ {6.55}\pm{0.25} $ & $ {12.12}\pm{0.46} $ & $ {0.973}\pm{0.029} $\\
 8.82  & $ {7.08}\pm{0.27} $ & $ {13.09}\pm{0.50} $ & 1\\
\hline
\end{tabular}
\end{center}
\end{table*}

\subsection{Electroluminescence gain measurements}
The magnitude of the S2 gain from electroluminescence in the gas region depends on the strength of the electric field in that region, as the field provides the energy needed for the electrons to excite gaseous xenon atoms and produce secondary light.


In order to properly disentangle the variation of the S2 signal size from electroluminescence gain and electron extraction efficiency from the liquid to the gas, the electroluminescence gain must be measured for each electric field and used to convert the S2 signal size into a number of detected electrons.  

\begin{figure}[ht]
\centering
\includegraphics[width=0.6\textwidth]{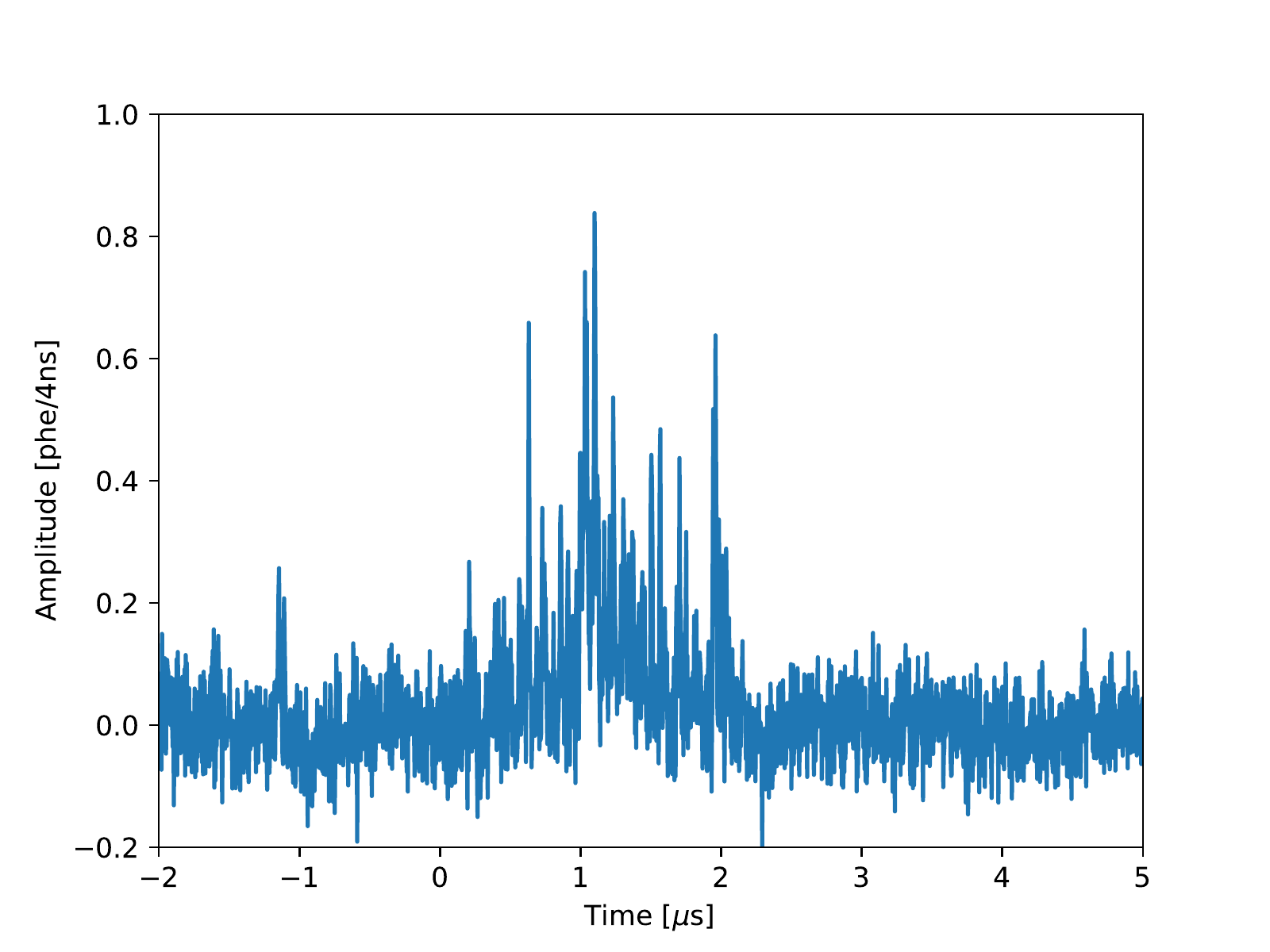}
\caption{Example signal waveform from a single extracted electron in the PIXeY detector.}
\label{fig:singleElectronWaveform}
\end{figure}

To determine the electroluminescence gain, the amount of S2 light produced by individual electrons drifting from the liquid surface to the anode was measured.  Single electrons are abundant in two-phase xenon detectors and have been studied previously in many different experiments~\cite{Edwards:2007, Santos:2011, Aprile:2014b}. An example single electron waveform is shown in figure~\ref{fig:singleElectronWaveform}.

\begin{figure}[ht]
\centering
\includegraphics[width=0.6\textwidth]{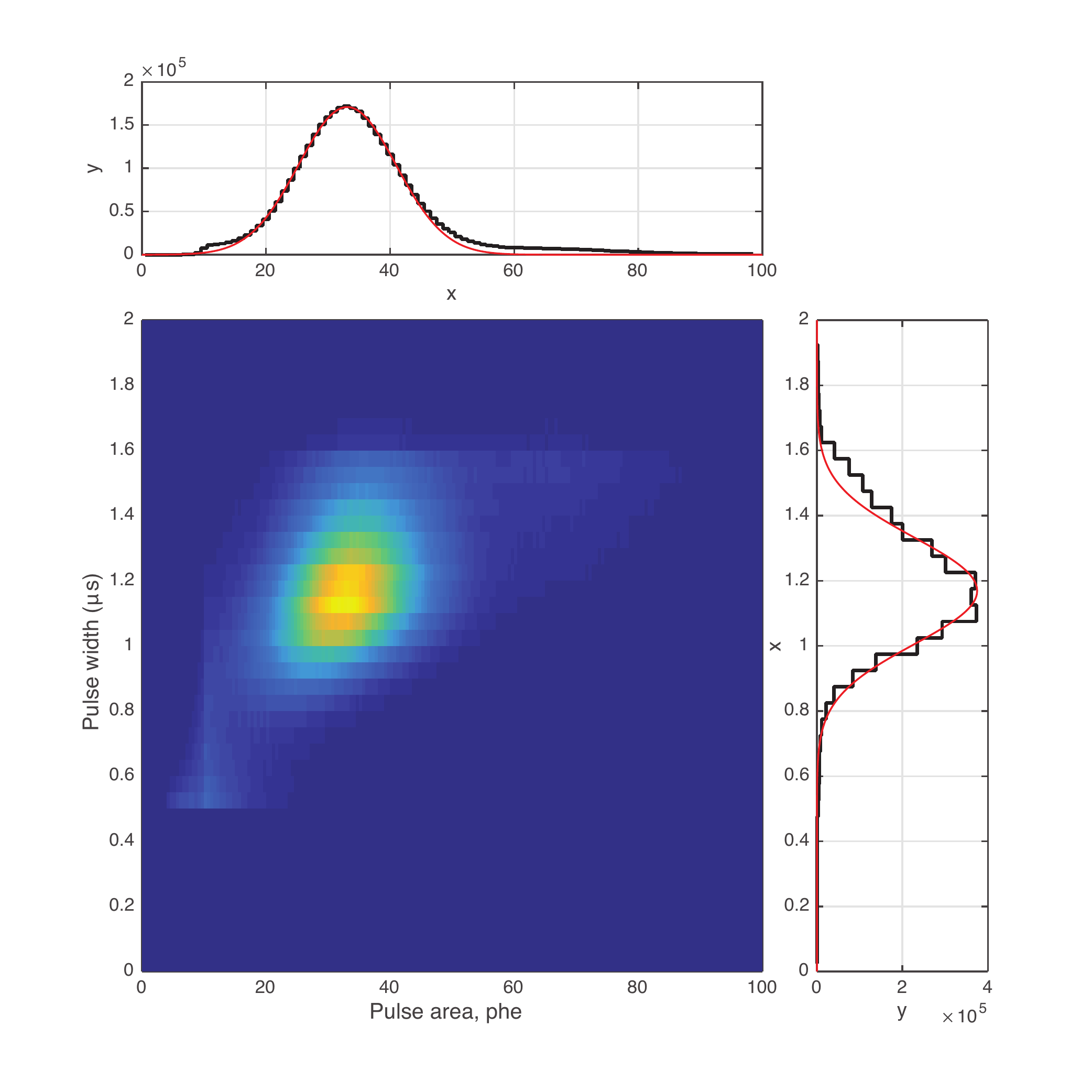}
\caption{Representative detector S2 response to single electrons extracted from the liquid into the gas.}
\label{fig:singleElectrons}
\end{figure}

\begin{figure}[!htb]
\centering
\includegraphics[width=0.6\textwidth]{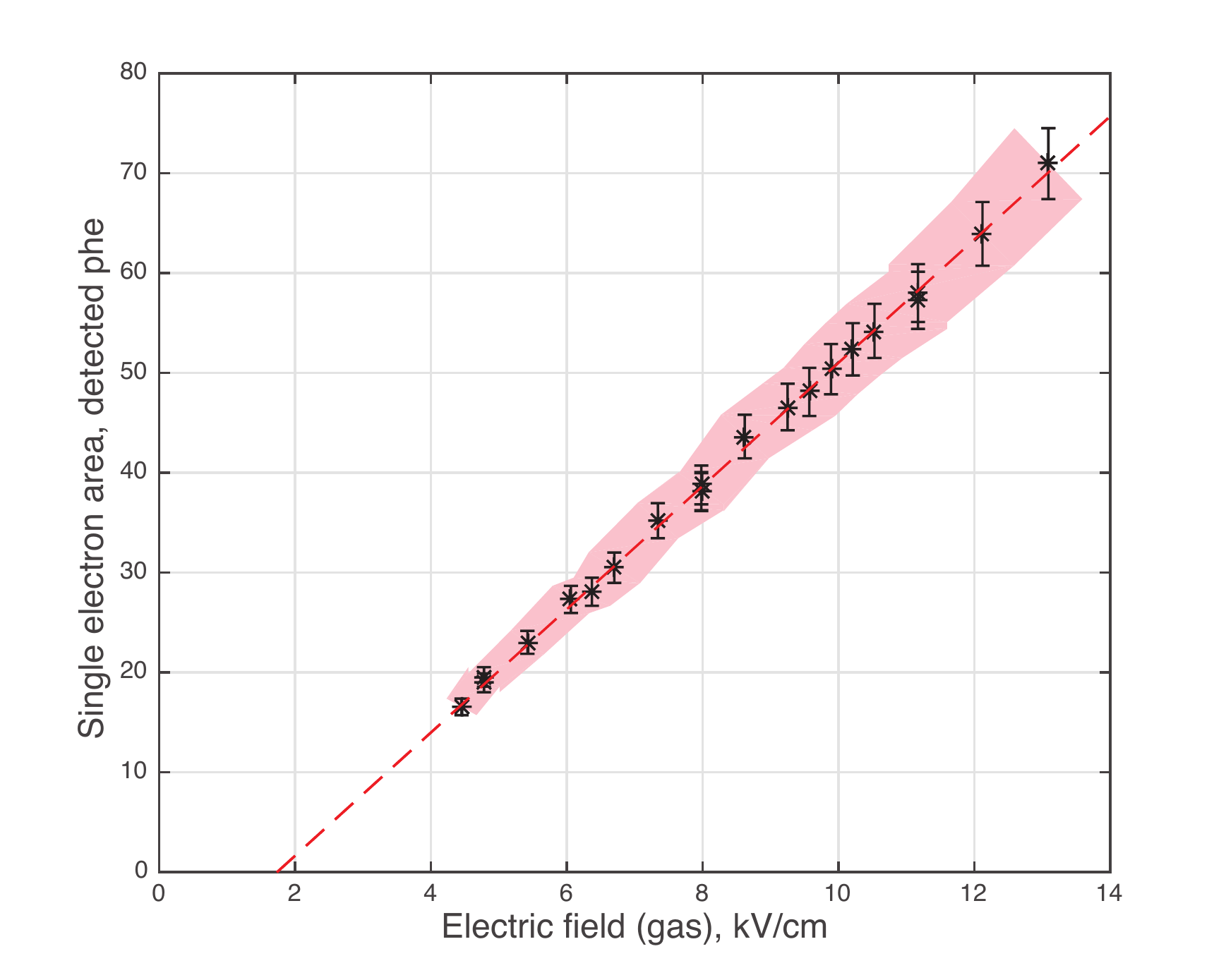}
\caption{Detected single photoelectrons per single electron as a function of electric field in the gas region, where the electric field is taken from the COMSOL field model. The red shaded region includes the uncertainty in electric field.}
\label{fig:singleElectronsField}
\end{figure}

Figure~\ref{fig:singleElectrons} shows an example of a single electron spectrum for one electric field configuration.  The single electron population is clearly defined in the pulse area against pulse width parameter space.  The width of the single electron signal is the time taken by the electron to cross the gas region from the liquid surface to the anode.  The single electron pulse area as a function of field is shown in figure~\ref{fig:singleElectronsField}.

\subsection{Extraction efficiency}
To measure the relative extraction efficiency, a source of consistent charge yield is required. The S2 yield of this source at a given extraction field may then be divided by the single electron S2 yield at the same field to determine the extraction efficiency. 

For the purposes of this study, two different mono-energetic calibration sources were used, \krm and $^{37}$Ar. As the drift field is kept constant throughout the experiment, the charge produced by each calibration source is also constant. The metastable \krm atom decays through two transitions, the first releasing 32.1~keV, followed by a second releasing 9.4~keV with a decay time constant of 154~ns. Each transition results mostly in conversion electrons. Occasionally 12 or 9.4 keV x-rays are emitted, but these have absorption lengths in liquid xenon of less than 10 $\rm \mu m$, much less than the electron diffusion ($\sim$ 1 mm) as the charge signal is drifted through the TPC. In addition, the 154~ns decay time constant between the 32.1 and 9.4~keV pulses is much less than the ~$\rm \mu s$ variation in electron drift time due to diffusion. As a result, the two interactions occur at essentially the same location and same time (effectively producing a single electron cloud) and S2 signals appear as a single pulse in the PMTs.  The \krm decay may then be treated as a mono-energetic S2 source of 42.5~keV.  $^{37}$Ar decays by electron capture through a number of transitions, with the dominant decay releasing 2.8~keV.  These are both low energy sources, with the $^{37}$Ar peak in the energy range of interest for WIMP search experiments. A description of the $^{37}$Ar source production and measurement with PIXeY may be found in \cite{Boulton:2017}. 

\begin{figure}[tb]
\centering
\includegraphics[width=0.6\textwidth]{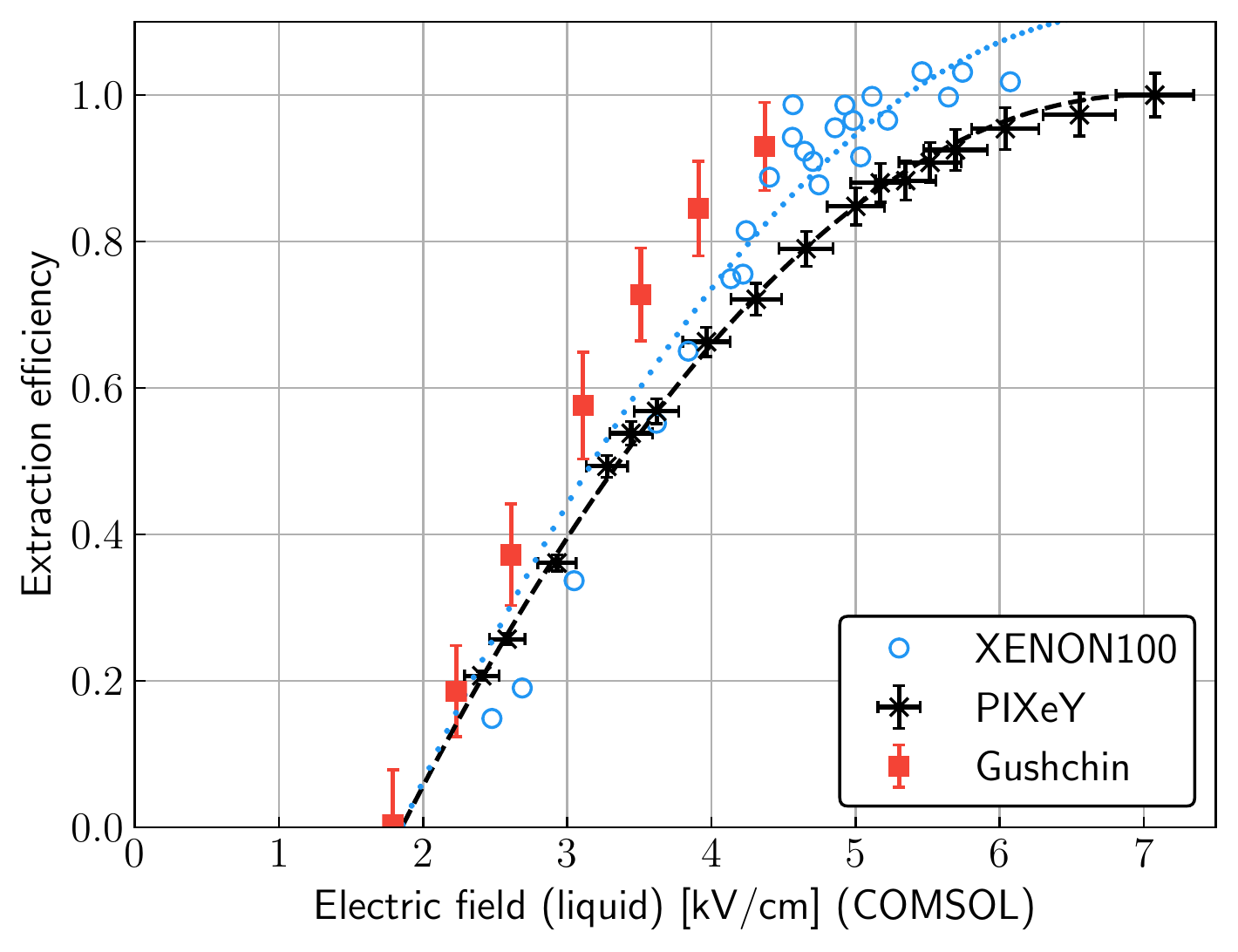}
\caption{Relative extraction efficiency as a function of electric field in the liquid xenon just below the liquid-gas surface, measured in PIXeY with mono-energetic peaks from \krm and $^{37}$Ar.  The electric field ($x$-axis) is calculated from a COMSOL electric field model. We compare the results with the absolute measurement by Gushchin~\textit{et al.}~\cite{Gushchin:1979} and the relative measurement by Aprile~\textit{et al.}~\cite{Aprile:2014b}. The black line is a best fit to the data, and the light blue dotted line is the same function multiplied by a constant value of 1.11519. Extraction fields in~\cite{Aprile:2014b} are quoted as fields in the gaseous xenon; here the extraction fields in~\cite{Aprile:2014b} are quoted as fields in the liquid xenon and divided by the dielectric constant of 1.85 for direct comparison with the present result.}
\label{fig:extractionEfficiency}
\end{figure}


Using these two sources, the size of the S2 signal is tracked at two different energies, thus reducing systematic uncertainties related to PMT saturation (which should be signal size dependent).  As with previous measurements in the literature, we normalize the maximum in the measured extraction (i.e. the number of S2 electrons at the highest fields) to 100\% electron emission from the liquid into the gas. 

Figure~\ref{fig:extractionEfficiency} shows the measured relative extraction efficiency as a function of electric field in the liquid.  A threshold is observed for the onset of extraction from the liquid phase into the gas, followed by an increase in extraction efficiency as extraction field is increased. We assign full extraction to correspond to our highest extraction field of 7.1 kV/cm in the liquid xenon.  Comparing the resulting efficiency to those previously seen in the literature, we measure a lower extraction efficiency over the full range of electric fields.  We show results from PIXeY plotted against the electric fields calculated using the COMSOL model.  The extraction efficiency is fit to a quadratic function, yielding a best fit of $y = - 0.03754x^2 + 0.52660x - 0.84645$, where $y$ is the extraction efficiency and $x$ is the electric field in the liquid xenon. The light blue line is the same function multiplied by a constant value of 1.11519. This suggests that previous studies, constrained to lower extraction fields, overestimated extraction efficiencies by approximately this factor.

The systematic error in the electric field applied, as described previously, is dominated by the geometry of the electric field region, with an overall systematic of about 5\%.  In the extraction efficiency, the dominant uncertainties arise from the uncertainty in the single electron signal size and that of the S2 peak position (for \krm and $^{37}$Ar). Combining these gives an overall error of 3\%, as shown in the right-most column of table~\ref{tab:fields}.  In addition, we believe any uncertainties due to PMT saturation are small due to the good agreement of the \krm and $^{37}$Ar signals, which give consistent extraction efficiencies while having an order of magnitude difference in signal amplitude.

\section{Summary}
The PIXeY detector has been used to probe a large range of extraction fields, allowing mapping of the electron extraction efficiency curve with better precision and over a wider range of fields than in previous experiments.  Due to its novel design features, PIXeY was able to apply electric fields across the liquid-gas interface up to 7.1~kV/cm (in the liquid).  Systematic errors in this measurement are constrained by the use of two radioactive sources with energies more than an order of magnitude apart. 

We observe extraction efficiencies that continue to increase at the highest extraction fields. This has the practical implication that additional charge signal may be attained through careful engineering of the gate-anode region, so as to enable a high extraction field.   Such measurements of the liquid xenon physics underlying two-phase detectors are important for both experimental design and interpretation of data from future large scale liquid xenon experiments. 

\begin{acknowledgments}

We acknowledge support from DHS grant 2011-DN-007-ARI056-02, NSF grant PHY-1312561, and DOE grant DE-FG02-94ER40870.  This research was conducted using computational resources and services of the Yale Science Research Software Core.  The $^{83}$Rb used in this research to produce \krm was supplied by the United States Department of Energy Office of Science by the Isotope Program in the Office of Nuclear Physics.

\end{acknowledgments}

\bibliographystyle{JHEP}
\bibliography{eee}

\end{document}